\documentclass[%
 reprint,
 amsmath,amssymb,
 aps,pra
]{revtex4-1}

\usepackage{float}
\usepackage{xcolor}

\usepackage{amsmath}
\usepackage{hyperref}
\usepackage{physics}
\usepackage{siunitx} 
\usepackage{mathtools}
\usepackage{braket}

\usepackage{float}
\usepackage{graphicx}
\usepackage{epstopdf}
\usepackage{dcolumn}
\usepackage{bm}

\newcommand{\Ca}{\ensuremath{^{40}\textrm{Ca}^+}}
\newcommand{\Ss}{\ensuremath{\textrm{S}_{1/2}}}
\newcommand{\Pd}{\ensuremath{\textrm{P}_{1/2}}}

\newcommand{\Dd}{\ensuremath{\textrm{D}_{3/2}}}

\newcommand{\DDi}{\ensuremath{\ket{\Dd,m_J\!=\!-3/2}}}
\newcommand{\SDi}{\ensuremath{\ket{\Ss,m_J\!=\!-1/2}}}
\newcommand{\DDf}{\ensuremath{\ket{\Dd,m_J\!=\!1/2}}}
\newcommand{\SDf}{\ensuremath{\ket{\Dd,m_J\!=\!3/2}}}
\newcommand{\DDfull}{\ensuremath{\DDi\!\rightarrow\!\DDf}}
\newcommand{\SDfull}{\ensuremath{\SDi\!\rightarrow\!\SDf}}
\newcommand{\SD}{\ensuremath{\Ss\rightarrow\Dd}}
\newcommand{\DD}{\ensuremath{\Dd\rightarrow\Dd}}
\newcommand{\Vsd}{50(2)\%}
\newcommand{\Vdd}{81(2)\%}
\newcommand{\Vsdphi}{54(2)\%}
\newcommand{\Vddphi}{89(2)\%}
\newcommand{\Vsdsim}{53.0\%}
\newcommand{\Vddsim}{92.2\%}

\newcommand{\reffig}[1]{Fig.~\ref{#1}}

\begin{document}

\preprint{APS/123-QED}

\title{Improving the Indistinguishability of Single Photons from an Ion-Cavity System}

\author{Thomas Walker}
\thanks{These authors contributed equally.}
\affiliation{Department of Physics and Astronomy, University of Sussex, Brighton BN1 9RH, United Kingdom}

\author{Samir Vartabi Kashanian}
\thanks{These authors contributed equally.}
\affiliation{Department of Physics and Astronomy, University of Sussex, Brighton BN1 9RH, United Kingdom}

\author{Travers Ward}
\affiliation{Department of Physics and Astronomy, University of Sussex, Brighton BN1 9RH, United Kingdom}

\author{Matthias Keller}
\email{M.K.Keller@sussex.ac.uk}
\affiliation{Department of Physics and Astronomy, University of Sussex, Brighton BN1 9RH, United Kingdom}


\begin{abstract}
We investigate two schemes for generating indistinguishable single photons, a key feature of quantum networks, from a trapped ion coupled to an optical cavity. Through selection of the initial state in a cavity-assisted Raman transition, we suppress the detrimental effects of spontaneous emission on the photon's coherence length, measuring a visibility of $\Vdd$ without subtraction of background counts in a Hong-Ou-Mandel interference measurement, the highest reported for an ion-cavity system. In comparison, a visbility of \Vsd{} was measured using a more conventional single photon scheme. We demonstrate through numerical analysis of the single photon generation process that the new scheme produces photons of a given indistinguishability with a greater efficiency than the conventional one. Single photon schemes such as the one demonstrated here have applications in distributed quantum computing and communications, which rely on high fidelity entanglement swapping and state transfer through indistinguishable single photons.
 \end{abstract}

\maketitle


\section{Introduction}
Entanglement between remote quantum systems is a prerequisite for distributed quantum computing \cite{moehring2007quantum,nickerson2014freely}, and quantum communication \cite{gisin2007quantum,simon2017towards}.
Various solid-state and atomic quantum systems have been proposed for this purpose, such as quantum dots \cite{gao2012observation,bussieres2014quantum,de2012quantum}, color centers in diamond (nitrogen-vacancy centers) \cite{togan2010quantum}, trapped neutral atoms \cite{volz2006observation,ritter2012elementary}, and trapped ions \cite{blinov2004observation,stute2012tunable}.
Coupling trapped ions to optical cavities combines the long trapping lifetimes and coherence times of ions \cite{wang2017single} with a highly controllable photonic interface and thus tunable temporal and spectral properties of emitted photons\cite{keller2004continuous,briegel1998quantum}.
Single photon emission with controlled frequency \cite{russo2009raman,keller2004calcium}, polarization \cite{barros2009deterministic}, and temporal shape \cite{keller2004continuous} has been demonstrated, as well as entanglement between ions and photons \cite{stute2012tunable}.
One established method for generating entanglement between remote quantum systems is to entangle each with a single photon as a flying qubit, and then project the stationary quantum systems into an entangled state through a Bell state measurement on the photons \cite{moehring2007quantum}.
The fidelity of the matter-matter entanglement process depends not only on the fidelity of the original ion-photon entanglements, but also the mutual distinguishability of the photons \cite{csimon2003robust,maunz2007quantum}. This distinguishability is reduced by experimental inhomogeneities and noise such as magnetic field strength or laser frequency jitter, and is ultimately limited by the atomic decoherence rates. In ion-cavity systems, ion-photon entanglement is generated through cavity-assisted Raman transitions, in which atomic population is transferred between two electronic states coupled by a laser and the cavity field, producing a photon in the cavity. The probability of spontaneous decay to the initial state means that one or more photons may scatter from the ion before the Raman process occurs, resulting in a time-jitter in the wavepacket of the photon emitted from the cavity. The observed photon is a probabilistic mixture of distinguishable time-shifted photons, each produced after a different number of scattering events, rather than a pure state \cite{ozeri2005hyperfine,barros2009deterministic}. Typically the initial state chosen is one with a high branching ratio from the excited state to increase efficiency and speed up state preparation, at the cost of the coherence of the generated photon. If instead the initial state were chosen with a low branching ratio, the trade-off would be reversed due to the low chance of decaying back to the initial state.

In this experiment we compare two cavity-assisted Raman schemes in trapped \Ca{}-ions with different initial states, shown in Figs. 1(b) and (c), by measuring the distinguishability of the emitted photons. In the first scheme, the initial and final states are Zeeman sublevels of the $4^2\Ss{}$ level and $3^3\Dd{}$ respectively, typical of experiments in \Ca{} and other species \cite{keller2004continuous,walker2018long,barros2009deterministic, stute2012tunable,crocker2019high}. In the second, the initial and final states are both within the $3^3D_{3/2}$ manifold. A similar scheme has previously been used in a free-space trapped-ion system for ion-photon state mapping \cite{eich2019single}. The lower decay rate from the excited state $4^2\Pd{}$ to \Dd{} ($\Gamma_{DP}=\SI{1.48}{\MHz}$) compared to \Ss{} ($\Gamma_{SP}=\SI{21.6}{\MHz}$) is expected to result in a significantly improved photon indistinguishability. To measure the distinguishability of the photons, the Hong-Ou-Mandel (HOM) two-photon interference effect is employed \cite{hong1987measurement}. First observed by Hong {\it{et. al.}} with photon pairs from parametric down conversion \cite{hong1987measurement}, the HOM effect has since been investigated in a wide variety of systems where it is commonly used to quantify photon distinguishability \cite{nisbet2011highly,gerber2009quantum,lang2013correlations,%
beugnon2006quantum,legero2003time,patel2010two,%
bernien2012two,eschner2001light}.

\begin{figure*}
    \centering
    \includegraphics[width=1.0\linewidth]{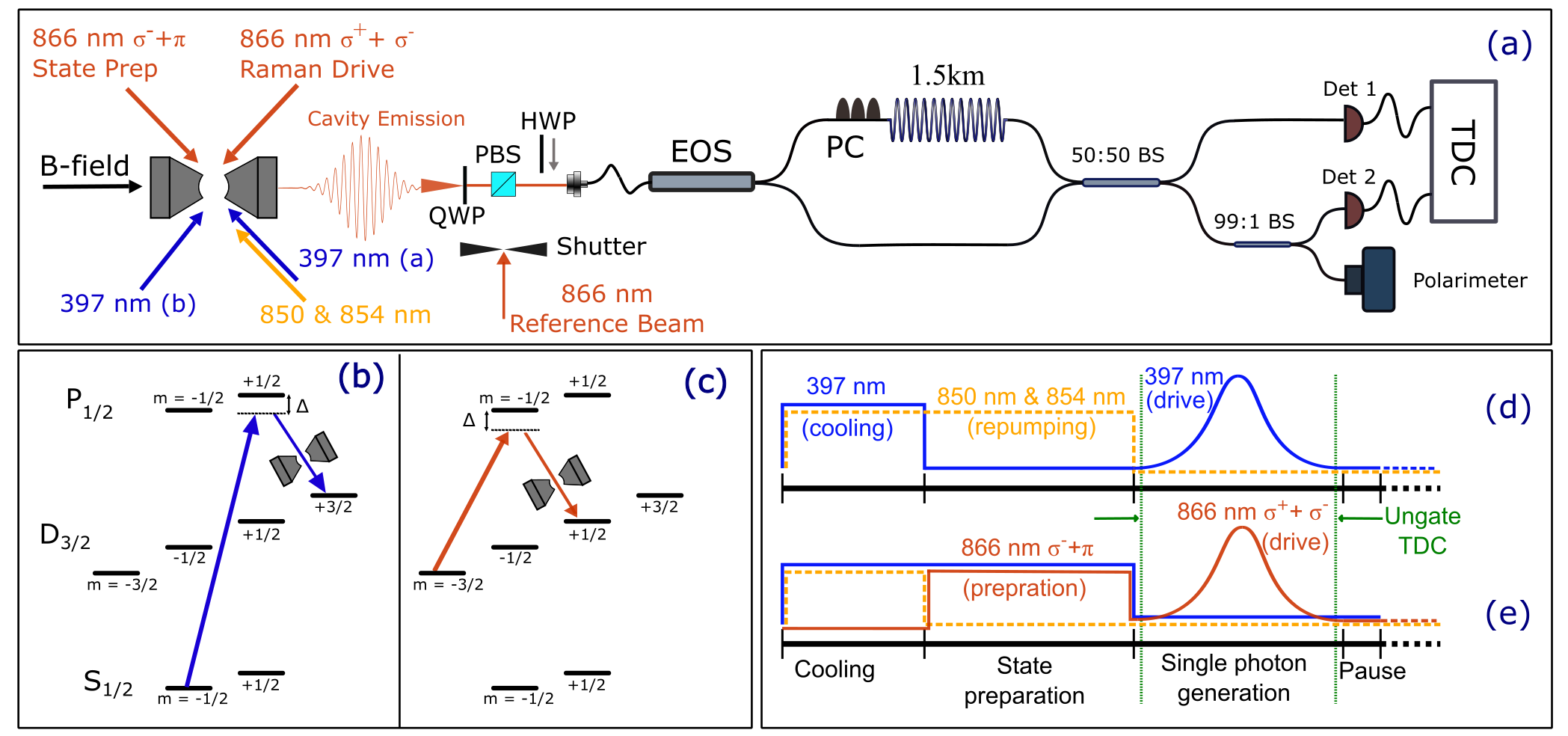}
    \caption{(a) The experimental setup. The lasers involved in this experiment and their orientations relative to the cavity axis are depicted. 
    A quarter-waveplate (QWP) and a polarising beam splitter (PBS) filter the cavity emission by polarisation. 
    An 866\,nm laser beam is optionally guided to the optical setup by releasing the shutter and moving a half-waveplate (HWP) into the beam path. The polarization in the delay arm may be adjusted using the polarization controller paddles (PC).
    The electro-optical switch (EOS) guides the photons alternately into the delay and the direct arms, such that successive photons meet at the 50:50 beam splitter (BS).
    A time-to-digital converter (TDC) records the photon detection times on the detectors Det 1 \& 2.
    (b,c) $\Ca$ level scheme showing two single photon schemes. $\Delta$ represents the detuning from atomic resonance of the laser and cavity. 
    (d,e) The sequence of laser pulses used to generate single photons.}
    \label{fig:setup}
\end{figure*}
\section{Setup}
The trap setup is shown in \reffig{fig:setup}(a). A single $^{40}$Ca$^+$ ion is trapped in a linear Paul trap. Four rf blade electrodes, with a distance of 475\,$\mu$m from the trap center to the tips of the electrodes, provide radial confinement with a secular frequency of 950\,kHz. Two dc end-cap electrodes separated by 5\,mm provide the axial confinement with a secular frequency of 900\,kHz. Stray electric fields that cause excess micromotion are compensated for by applying dc voltages onto the rf electrodes. A pair of highly reflective mirrors embedded in the end-caps form an optical cavity along the trap axis. The mirrors are shielded by the end-cap electrodes, avoiding distortion of the trapping potential caused by the dielectric surfaces of the mirrors. The cavity length is 5.75\,mm, with mirror transmissivities of 100 and 5\,ppm at 866\,nm, and radii of curvature of 25.4\,mm, leading to a cavity finesse of ~60,000 and an ion-cavity coupling strength $g_0 = 2\pi \times 0.8$\,MHz. Three Helmholtz coils located around the trap produce a magnetic field to align the quantization axis co-linear to the cavity axis and to split the Zeeman sublevels.

The laser beams for cooling, state preparation, pumping and re-pumping are injected into the system through the gaps between the rf electrodes.
In order to measure the indistinguishability of the photons emitted by the system through HOM interference, two photons must arrive at the same time at a 50:50 beam splitter. To this end, two consecutive photons are generated. The first photon is delayed by an optical delay fiber to arrive in coincidence with the second photon at the beam splitter.
The HOM interference setup is shown in Fig. \ref{fig:setup}(a). 
The cavity emission first passes through a quarter-waveplate and a polarising beam splitter cube (PBS), to clean the photon's polarisation. It then passes through two shortpass filters to remove the cavity locking light. 
The filtered emission is then coupled into a single-mode fiber-based electro-optical switch (EOS) (Agiltron, NanoSpeed Ultra-Fast). The input light is directed down one of two output ports. 
One output of the EOS leads directly to the 50:50 fiber-based beam splitter (FBS), while the other is connected to the FBS via the 1.5\,km delay line fiber. The two output ports of the FBS then lead to separate superconducting-nanowire single photon detectors (SSPDs) (Photonspot inc.) with a rated quantum efficiency at 850\,nm of 80\%. 
A time-to-digital converter (TDC) (quTAU, qutools) records timestamps for each detector click. 
To measure the polarisation distortion caused by the birefringence of the delay line fiber, a 99:1 beam splitter taps off 1\% of the signal from one of the FBS outputs to a polarimeter. This distortion may then be corrected for using polarisation control paddles.
As the polarimeter is not sensitive enough to measure single photons, an 866\,nm laser is overlapped with the cavity emission at the PBS, and has its polarization matched to the cavity emission with a half-waveplate. This beam is blocked by a shutter during data collection, and the experiment is paused regularly to correct for the polarization distortion.
To avoid coupling losses, all fiber-fiber connections in this setup are spliced.
To ensure consecutive photons arrive at the same time at the FBS, the single photon sequence is repeated with a period equal to the travel time of light through the delay line fiber ($7.38\,\mu$s), with the EOS output re-directing the cavity emission between cycles.
\section{Results}
The single photons are generated in the sequence depicted in \reffig{fig:setup}(d) and (e).
The ion is first Doppler cooled by a 397\,nm beam for \SI{1.5}{\micro\second} while lasers at 850 and 854\,nm repump the ion from the metastable states back into the cooling cycle.
To generate a single photon from the initial state $\Ss$ the 397\,nm laser is switched off for 2.5\,$\mu$s to state prepare the ion (\reffig{fig:setup}(d)).
The single photon is then generated through a cavity-assisted Raman transition between the $\Ss,m=-1/2$ and $\Dd,m = 3/2$ states using a 397\,nm laser beam with $\sigma^+$ and $\pi$ polarization for \SI{2.5}{\micro\second} (see Fig. \ref{fig:setup}(d)).
The intensity of this laser has a Gaussian temporal shape with a width of 450\,ns and amplitude of $\Omega=2\pi\times\SI{11}{\MHz}$ and is red-detuned \SI{24}{\MHz} from resonance.
The long lifetime of the $\Dd$ metastable state guaranties the creation of no more than a single photon in the cavity at one time.
The splitting of the Zeeman sublevels by the magnetic field is much larger than the linewidth of the transition, allowing the selection of a specific Raman transition.
A delay of a few 100\,ns before and after the photon generation step ensures the complete switch-off of all the other lasers and decay of the cavity population to avoid the creation of multiple photons.

The overall probability of generating and detecting these single photons was $P_{\rm det, SD} = \SI{0.36 \pm 0.0003}{\percent}$. From this and the known system losses we estimate a probability of emitting a photon from the cavity of $P_{\rm emit, SD} = \SI{1.81}{\percent}$. Numerical simulations of the system give an expected emission efficiency of $P_{\rm emit, SD} = \SI{1.8}{\percent}$.
A Hanbury-Brown-Twiss (HBT) measurement confirms the system is a single photon source with $g^{(2)}(0)=0.0017(12)$ \cite{walker2018long}.
The TDC records the arrival time of the single photons on two channels only within the single photon generation part of the sequence.
An additional electronic pulse is generated every 256 cycles and time-stamped by another channel of the TDC, synchronising the experimental sequence and the TDC to obtain the temporal profile of the single photons (\reffig{fig:phot_shape}). 

\begin{figure}[h]
    \includegraphics[width=\linewidth]{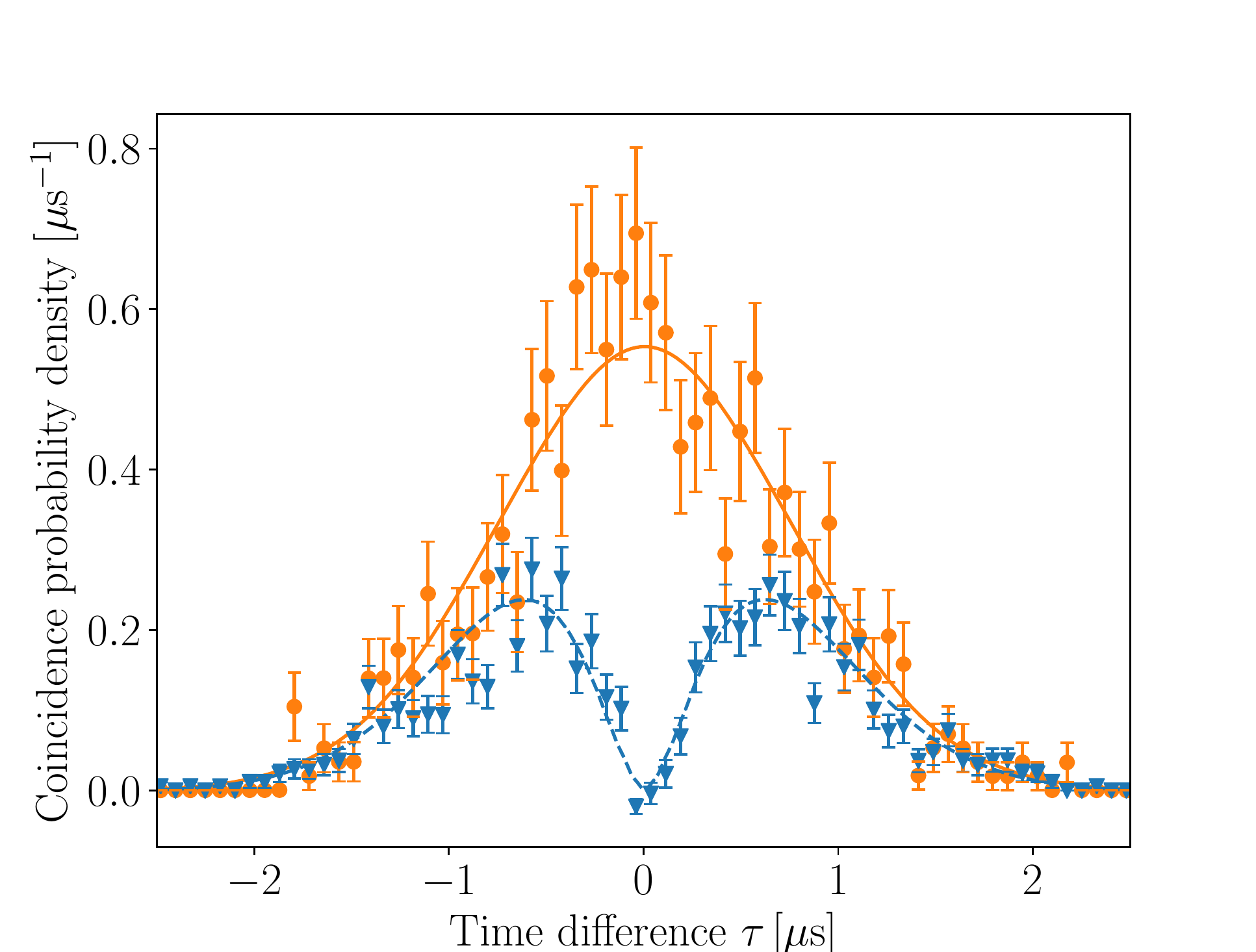}
    \caption{Coincidence probability density for single photons produced using the $\SD$ scheme, normalising the area under the curve for perpendicular polarized photons to unity. The orange circles show the reference signal obtained by fully distinguishable photons with orthogonal polarization, while the blue triangles show the coincidences between two detectors with parallel polarized photons. Error bars representing one standard deviation. The solid orange and dashed blue lines show the expected coincidence probability from numerical simulation for the orthogonal and parallel cases.}
    \label{fig:hom_S2D}
\end{figure}

\begin{figure}[h]
\includegraphics[width=\linewidth]{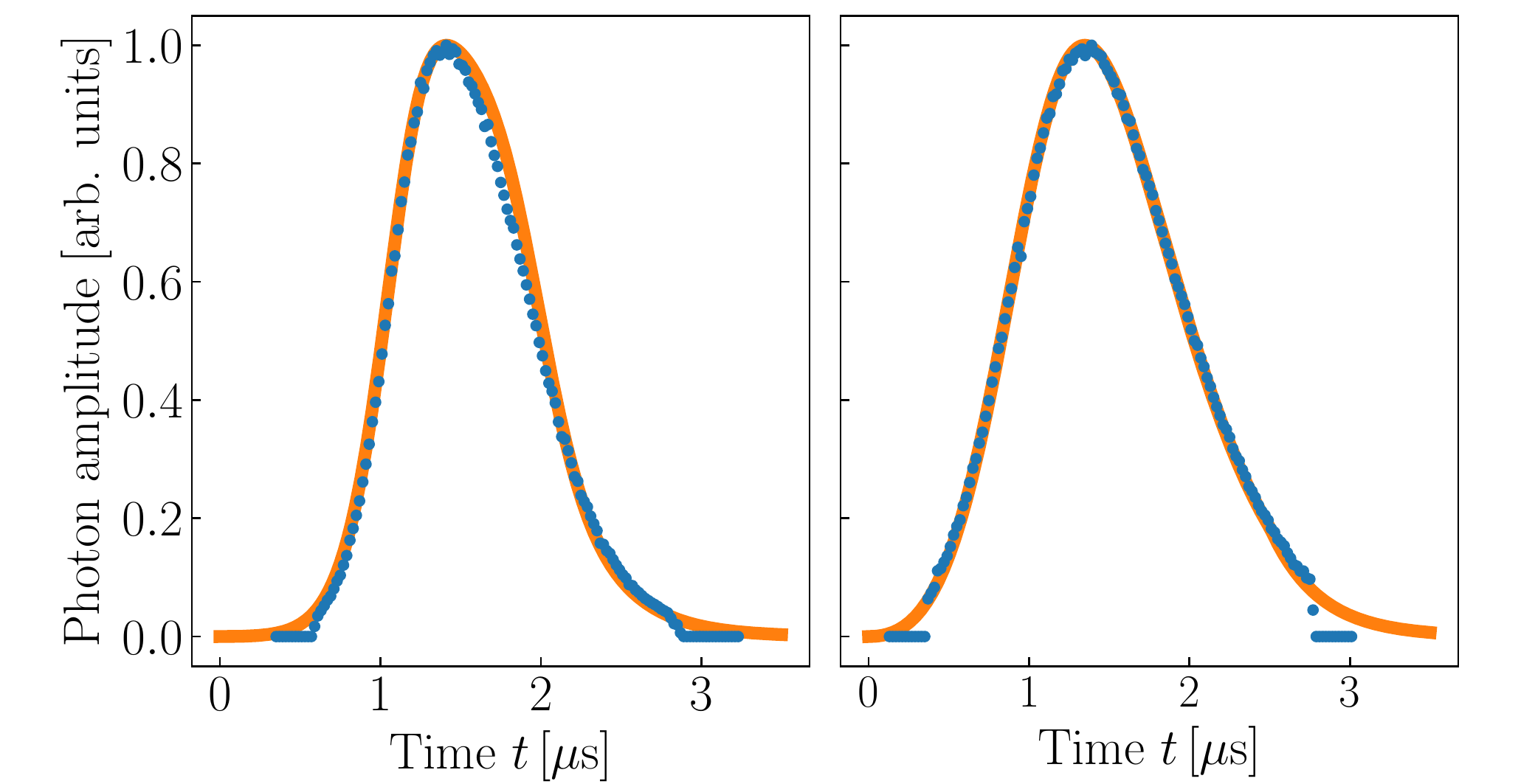}
    \caption{Temporal probability distribution of detecting single photon shown as blue dots for the $\SD$ scheme (left) and $\DD$ (right). 
    To extract this plot, all the photon arrival times with respect to the sequence trigger during the measurements are sorted into 20\,ns time bins and the resulting histograms are normalized to unity. 
    The solid lines are obtained by numerical simulation.}
    \label{fig:phot_shape}
\end{figure}

Cross-correlation between two detectors is obtained and plotted in a histogram with a bin-width of \SI{75}{\ns} (\reffig{fig:hom_S2D}).
The clear dip at the center is characteristic of a HOM interference pattern with partially distinguishable photons \cite{nisbet2011highly}. 
As a reference measurement, the experiment is repeated with fully distinguishable photons by rotating the polarization of photons through the delay fiber perpendicular to those from the direct path using the polarization controller.

The HOM visibility is defined as \cite{legero2003time}
\begin{equation}
    \textrm{V} = 1- \frac{\int_{\scriptscriptstyle{-T/2}}^{\scriptscriptstyle{T/2}} C_{\parallel}(\tau) d\tau}{\int_{\scriptscriptstyle{-T/2}}^{\scriptscriptstyle{T/2}} C_{\bot}(\tau) d\tau},
    \label{eq:vis}
\end{equation}
where $C_{\parallel,\bot}(\tau)$ are number of coincidences for parallel and perpendicular polarization respectively, and $T$ is equal to twice the single photon window.
2,788,867 photons were detected in order to measure the HOM interference signal, and 1,511,965 photons to make the fully distinguishable (perpendicular polarization) histogram. We obtain a visibility $V_{\rm SD} = \Vsd$.
The effect of the distinguishability caused by scattering on the $\Pd\rightarrow\Ss$ transition can be seen as wings about $\tau=0$ in the HOM histogram.
The system is simulated through numerical solutions of the master equation for an 8-level ion coupled to a cavity using QuTiP \cite{johansson2013qutip}. To obtain the expected HOM interference pattern, we calculate the first- and second-order coherence functions of the cavity emission \cite{fischer2016dynamical}. The simulations are scaled to the data in \reffig{fig:hom_S2D} by normalizing the area under the curves for perpendicular polarization to unity. A degree of distinguishability due to polarisation drift in the delay fiber is expected. This polarisation mode mismatch is incorporated in the simulations and fitted to the data and then subtracted. An average angle offset of $\phi=\ang{12}$ was found. Accounting for this, a visibility of \Vsdphi{} is extracted, in agreement with the simulated value $V_{\rm sim, SD} = \Vsdsim{}$.
The temporal profile of the cavity emission is shown in \reffig{fig:phot_shape}(a) together with the simulated profile, which show good agreement.

The second scheme, \DD{}, uses the same sequence structure and timing but different lasers (see \reffig{fig:setup}(c)).
After Doppler cooling as in the previous scheme, the ion is prepared into the \DDi{} state by optically pumping with a 397\,nm beam and an 866\,nm beam, both polarized $\sigma^- + \pi$.
The Raman transition from \DDi{} to \DDf{} is then driven by a $\sigma^+ + \sigma^-$-polarized, \SI{24}{\MHz} blue-detuned \SI{866}{\nm} pulse of amplitude $\Omega=2\pi\times\SI{5.5}{\MHz}$.
This has a total efficiency of $P_{\rm det, DD} = \SI{0.05904 \pm 0.00003}{\percent}$, giving an emission probability of $P_{\rm emit, DD}=\SI{0.27}{\percent}$. This is lower than the value of $P_{\rm emit, DD}=\SI{0.75}{\percent}$ from simulations, likely due to the state preparation efficiency, which is limited by the time available for state preparation and polarisation purity of the lasers. The temporal profile of the cavity emission is shown in figure Fig \ref{fig:phot_shape}(b) together with a simulated profile.
A HBT measurement gives $g^{(2)}(0)=0.036(16)$. The small offset from zero is due to both the background rate and chance of two-photon events caused by $\sigma^-$ component of the polarization of the Raman drive beam, present because of the geometry of the setup. 

\begin{figure}[h]
    \includegraphics[width=\linewidth]{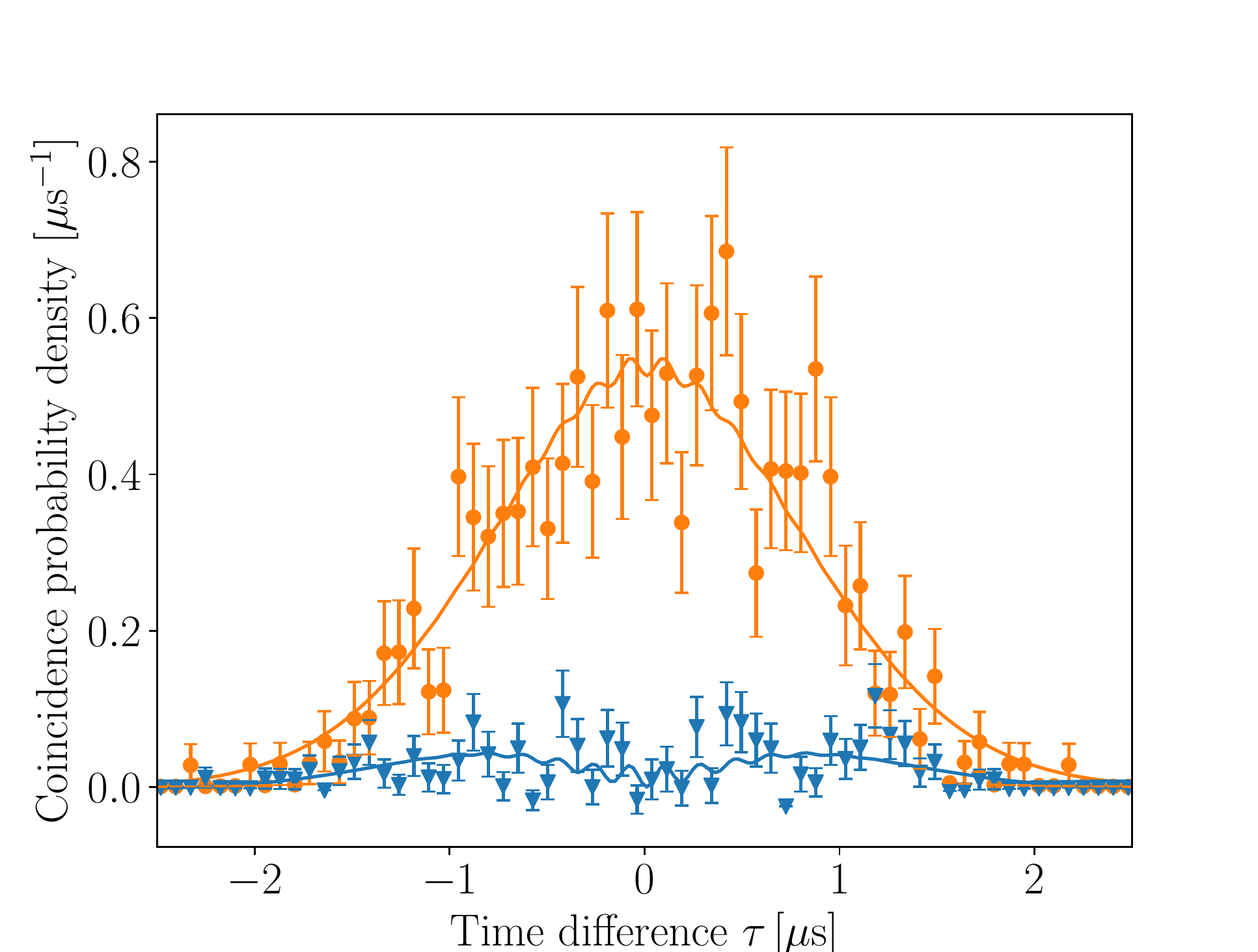}
    \caption{Time-resolved HOM interference signal for single photons produced using the $\DD$ scheme. As in Fig. \ref{fig:hom_S2D}, the orange dots and blue triangles show the signal obtained from orthogonal and parallel polarized photons, respectively. The solid and dashed lines are from numerical simulations of the system.}
    \label{fig:hom_D2D}
\end{figure}

\reffig{fig:hom_D2D} shows the time-resolved HOM signal and comparison with the reference measurement together with the simulated HOM interference. To create this plot $4,762,676$ photons with parallel polarization were collected along with $4,273,969$ photons with perpendicular polarization. The oscillations visible on the simulated curve are due to the $\sigma^-$ polarization component of the Raman laser, which couples back to the initial state, resulting in a beat note. A visibility of \Vdd{} is extracted directly from the data, increasing to \Vddphi{} when accounting for the polarization mismatch. The visibility from simulations $V_{\rm sim, DD} = \Vddsim{}$ is slightly higher than that from the data, as with the \SD{} case.

\section{Comparison}

There is a clear improvement in the indistinguishability of the single photons produced in our ion-cavity system by choosing $\Dd$ as the initial state over $\Ss$. Simulations show that a HOM visibility of $V = \Vddsim$ for the $\DD$ scheme is achievable, with the measured value limited by polarisation drift. The higher efficiency of the \SD{} scheme is not primarily due to the population recycling effect, but due to the higher state preparation and more favourable Clebsch-Gordan coefficienct for the transition. A better comparison would use the Raman transition $\ket{\Dd,m_J\!=\!-1/2}\rightarrow\ket{\Dd,m_J\!=\!3/2}$, which shares the Clebsch-Gordan coefficient with \SDfull{}. This was not feasible in this experiment, as the time available for state preparation was limited by the sequence repetition rate, fixed by the length of the delay fiber. This limit is therefore only a factor of the measurement setup used here and would not exist in a practical quantum network. Near-unity-fidelity state preparation has been demonstrated in \Ca{}, taking around \SI{10}{\micro\second} \cite{sorensen2006efficient}. Simulations indicate that in this case the relative efficiency difference reduces to a few percent, with little change in the HOM visibilities.

\begin{figure*}[ht]
\centering
\includegraphics[width=\linewidth]{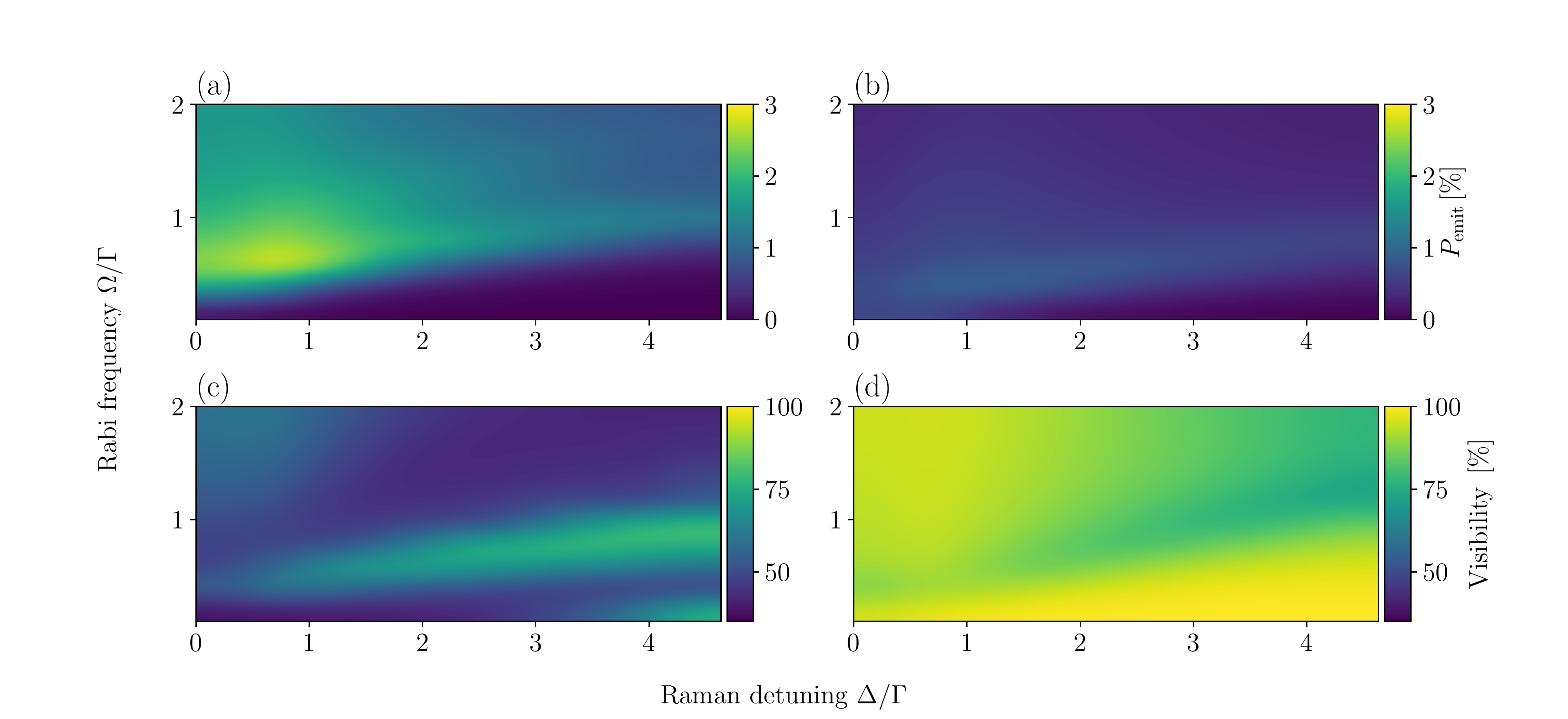}
\caption[Visibility and efficiency heat maps.]{Simulated visibility and and emission probability for the \SD{} and \DD{} schemes plotted against Raman detuning and drive laser Rabi frequency in units of the transition linewidth $\Gamma_{SD}$. \textbf{(a)} \SD{} efficiency. \textbf{(b)} \DD{} efficiency. \textbf{(c)} \SD{} visibility. \textbf{(d)}  \DD{} visibility.}
\label{fig:heat}
\end{figure*}

There are ways to improve the indistinguishability of photons, which we will now consider.  The experimental Raman laser parameters (peak Rabi frequency $\Omega$ and detuning $\Delta$) were chosen to maximise the photon generation efficiency in each case. However, both the probability of spontaneous decay and the photon generation efficiency depend nonlinearly on the laser parameters. It is therefore possible that the \SD{} scheme could be superior for a particular set of powers and detunings. To investigate this, the HOM visibility and emission probability were simulated over a range of powers and detunings, shown in \reffig{fig:heat}. Over the range of values considered, any desired visibility can be achieved with a greater efficiency in the \DD{} scheme than \SD{}.

\begin{figure}[ht]
\centering
\includegraphics[width=\linewidth]{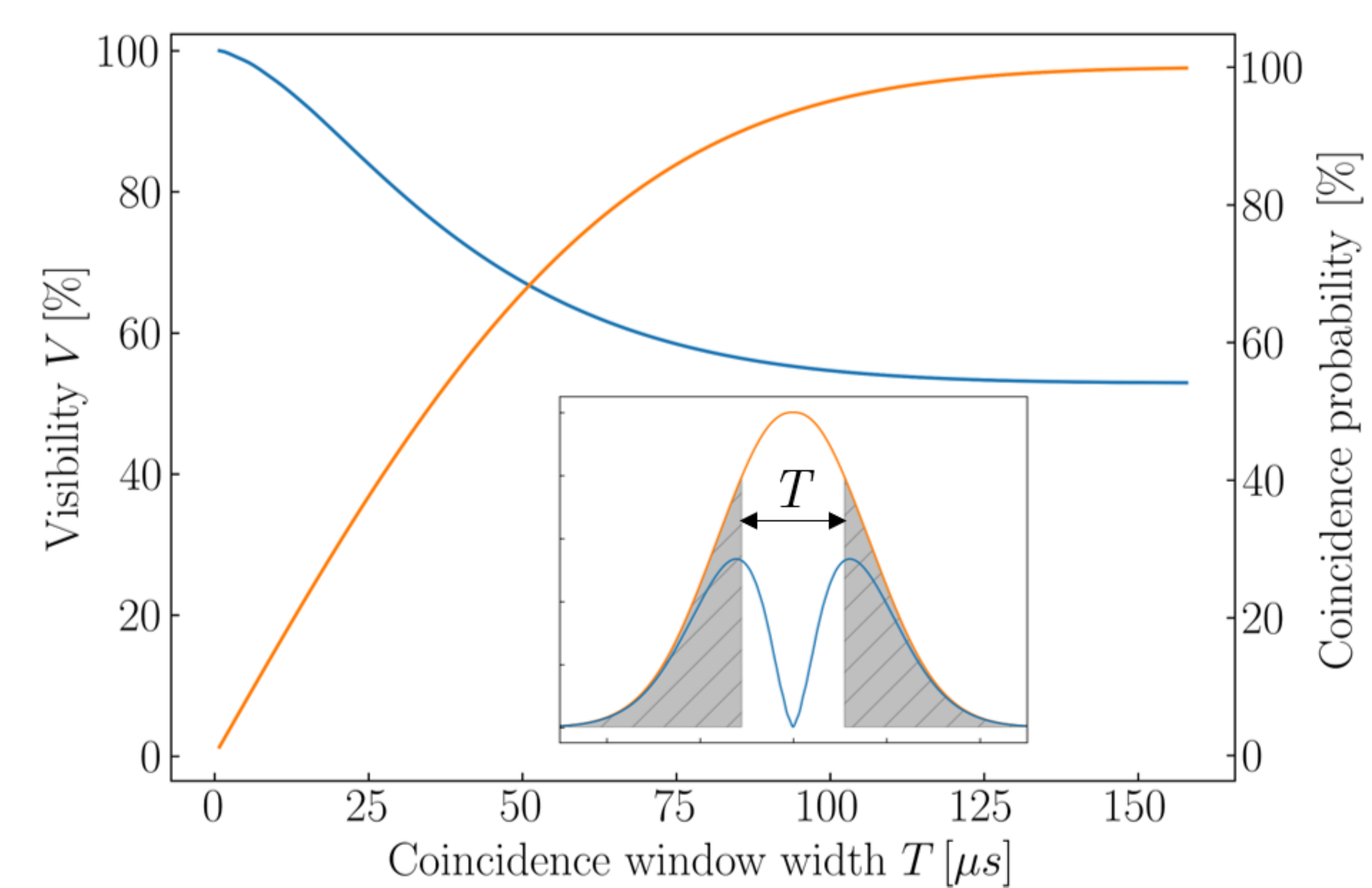}
\caption[Coincidence rate and visibility versus window size.]{Visibility (blue) and coincidence count probability (orange) with varying window size $T$ for the simulation of the \SD{} scheme. The coincidence count probability is normalised to the experimental rate. The insert demonstrates the temporal filtering; coincidences in the grey shaded area are neglected, increasing the visibility.}
\label{fig:window}
\end{figure}

Due to the shape of the HOM interference pattern, it is possible to increase the effective visibility by temporally filtering the coincidence counts. \reffig{fig:window} shows how the visibility and rate of coincidence counts for perpendicular polarisation changes with the maximum time between photon counts in the \SD{} scheme. For a coincidence window width giving a greater visibility in the \SD{} scheme, the coincidence count rate would always be lower than that of the \DD{} scheme. 

\section{Conclusion}
We have investigated two schemes to generate single photons from an ion-cavity system and measured the emission probability and indistinguishability of the photons. While the photon generation efficiency from the cavity-assisted Raman transition \SDfull{} was greater, the indistinguishability of the emitted photons is reduced by multiple scattering events on the $\Pd \rightarrow \Ss$ transition during the photon generation. Using a Raman \DDfull{} transition significantly improved the indistinguishability of the produced photons by reducing the probability of these scattering events occurring. Numerical analysis shows that measures to improve the visibility of the $\SD$ scheme tend to lower the effective efficiency compared to the $\DD{}$ scheme. This also holds in the strong coupling regime. Further, with better state preparation techniques, the difference in the efficiencies of the schemes could be greatly reduced. For applications which require indistinguishable photons, including probabilistic entanglement schemes, quantum information processing and quantum key distribution, it is advantageous to select a single photon generation scheme which limits the effects of spontaneous emission, such as the one demonstrated here.  Based on the coincident counts from our system, the expected entanglement rate between two identical systems is similar to the first free space demonstration of probabilistic entablement with trapped ions \cite{moehring2007quantum}. Employing our scheme in a strongly coupled system such as \cite{takahashi2020strong}, the entanglement rate increases well beyond the latest free space demonstrations \cite{stephenson2020high}. This scheme also provides a path to time-bin encoding in ion-cavity systems, where the coherence of the photon must be preserved across multiple time-bins.


\begin{acknowledgments}
We gratefully acknowledge support from EPSRC through the UK Quantum Technology Hub: NQIT - Networked Quantum Information Technologies EP/M013243/1.
\end{acknowledgments}


\bibliography{HOM_bib}

\end{document}